# Coefficient Decomposition of Spatial Regressive Models Based on Standardized Variables

Yanguang Chen

(Department of Geography, College of Urban and Environmental Sciences, Peking University, 100871, Beijing, China. Email: chenyg@pku.edu.cn)

**Abstract**: Spatial autocorrelation analysis is the basis for spatial autoregressive modeling. However, the relationships between spatial correlation coefficients and spatial regression models are not yet well clarified. The paper is devoted to explore the deep structure of spatial regression coefficients. By means of mathematical reasoning, a pair of formulae of canonical spatial regression coefficients are derived from a general spatial regression model based on standardized variables. The spatial auto- and lag-regression coefficients are reduced to a series of statistic parameters and measurements, including conventional regressive coefficient, Pearson correlation coefficient, Moran's indexes, spatial cross-correlation coefficients, and the variance of prediction residuals. The formulae show determinate inherent relationships between spatial correlation coefficients and spatial regression coefficients. New finding is as below: the spatial autoregressive coefficient mainly depends on the Moran's index of the independent variable, while the spatial lag-regressive coefficient chiefly depends on the cross-correlation coefficient of independent variable and dependent variable. The observational data of an urban system in Beijing, Tianjin, and Hebei region of China were employed to verify the newly derived formulae, and the results are satisfying. The new formulae and their variates are helpful for understand spatial regression models from the perspective of spatial correlation and can be used to assist spatial regression modeling.

# 1 Introduction

One of basic conditions for effective linear regressive modeling is that there is no significant correlation among sampling points. For spatial samples in a geographical region, the necessary conditions for building linear functional relation between independent variable $x$ and dependent variable $y$ through regressive analysis can be outlined as follows. First, there is significant correlation between $x$ and $y$; Second, variables $x$ and $y$ bear no significant spatial autocorrelation; Third, there is no significant spatial cross-correlation between $x$ and $y$. If the above conditions are met and the samples are large enough, we can establish a linear regression model or a simple nonlinear model. The nonlinear model is able to be linearized by mathematical transformation. For other disciplines such as biology and ecology, researchers can control the significance of spatial autocorrelation of samples by adjusting the distance between sampling points. Unfortunately, in most majority of cases, the spatial proximity of geographical systems such as network of cities cannot be changed. In this instance, one of approaches to make effective regressive models it to take spatial autoregressive term or spatial lag-regressive term into account. Therefore, geo-spatial autoregressive modeling method came into being.

In statistical analysis, correlation and regression often appear as a pair of twin concepts. However, for spatial statistics and modeling, we have to know clear mathematical relationships between spatial correlation coefficients and spatial autoregressive coefficients. Spatial correlation coefficients involve both spatial autocorrelation coefficient, that is, Moran's index, and spatial cross-correlation coefficient. Spatial cross-correlation coefficient sometimes was treated as two-variable autocorrelation coefficient in literature, but this understanding is inaccurate. Autocorrelation reflects intra-sample correlation, while cross-correlation reflects inter-sample correlation (Chen, 2015). Standardizing variables by means of $z$-score, we can express Moran's index and spatial cross-correlation coefficient as quadratic form. Spatial autoregressive models can also be simplified in form by using standardized variables. Due to the inherent relationships between correlation and regression, maybe we can reveal the deep structure of spatial autoregressive coefficients and thus reveal the internal relationships between spatial correlation and spatial auto-regression. This paper is devoted to derive the mathematical expressions of spatial autoregressive coefficients based on spatial auto- and cross-correlation coefficients and related statistics and parameters. The other parts are arranged as follows. In Section 2, the quadratic forms of Moran's index and spatial cross-correlation correlations are reviewed, and new formulae of spatial autoregressive coefficients are derived using the common methods from linear algebra. In Section 3, based on the urban systems in Beijing-Tianjin-Hebei region of China, empirical analyses are made to verify the newly derived formulae. In Section 4, several questions are discussed, and finally, in Section 5, the discussion is concluded by summarizing the main points of this study.

## 2 Models

### 2.1 Reviewing basic measurements and relations

Correlation and regression represent two different sides of the same coin in statistics. Similarly, auto-correlation is associated with auto-regressive modeling. To derive new formulae of spatial autoregressive model parameters, it is necessary to review a number of formulae for spatial autocorrelation and spatial cross-correlation as well as several conventional statistic measurements. For simplicity, all the basic variables are standardized by means of z-score. Based on standardized variable, spatial correlation coefficients and spatial regression models can be expressed in simple and canonical forms. And thus, the mathematical reasoning processes will become clear and succinct. In this way, we can reveal many hidden relationships unknown previously. Suppose there are $n$ elements (e.g., cities) in a system (e.g., a network of cities) which can be measured by two standardized variables (e.g., city size and urban area), $x$ and $y$. A pair of standardized vectors can be expressed as below:

$$\mathbf{x} = \begin{bmatrix} x_1 & x_2 & \cdots & x_n \end{bmatrix}^T, \tag{1}$$

$$\mathbf{y} = \begin{bmatrix} y_1 & y_2 & \cdots & y_n \end{bmatrix}^T, \tag{2}$$

where $x_i$ and $y_i$ are two size measurements of the $i$th observation ($i=1,2,\ldots,n$), the superscript "T" implies matrix transposition. According to the $z$-score properties, the means of the two variables are 0, and the standard deviations of them are 1. The inner products of the two variables are $n$, that is

$$\mathbf{x}^T\mathbf{x} = \mathbf{y}^T\mathbf{y} = n. \tag{3}$$

The coefficient of Pearson correlation between $x$ and $y$ can be expressed as

$$R = \frac{1}{n}\mathbf{x}^T\mathbf{y} = \frac{1}{n}\mathbf{y}^T\mathbf{x}, \tag{4}$$

where $R$ denotes the product-moment correlation coefficient. Spatial autocorrelation coefficient, i.e., Moran's indexes, of the two measures are as follows (Chen, 2013)

$$I_x = \mathbf{x}^T\mathbf{W}\mathbf{x}, \tag{5}$$

$$I_y = \mathbf{y}^T\mathbf{W}\mathbf{y}, \tag{6}$$

where $\mathbf{W}=[w_{ij}]$ denotes the global normalized spatial weight matrix. The three properties of the matrix are as follows: (1) Symmetry, i.e., $w_{ij}=w_{ji}$; (2) Zero diagonal elements, namely, $|w_{ii}|=0$, which implies that the entries in the diagonal are all 0; (3) Global normalization condition, that is, $\sum_i\sum_j w_{ij}=1$. The spatial cross-correlation coefficient between the two variables are as below (Chen, 2015):

$$R_{xy} = \mathbf{x}^T\mathbf{W}\mathbf{y} = \mathbf{y}^T\mathbf{W}\mathbf{x} = R_{yx}. \tag{7}$$

All the correlation coefficients can be unified in a logic framework. Define a matrix as follows

$$\mathbf{X} = \begin{bmatrix} \mathbf{x} & \mathbf{y} \end{bmatrix}. \tag{8}$$

Thus a 2-dimensional Pearson correlation matrix $\mathbf{R}$ can be expressed as

$$\mathbf{R} = \frac{1}{n}\mathbf{X}^T\mathbf{X} = \frac{1}{n}\begin{bmatrix} \mathbf{x}^T \\ \mathbf{y}^T \end{bmatrix}\begin{bmatrix} \mathbf{x} & \mathbf{y} \end{bmatrix} = \frac{1}{n}\begin{bmatrix} \mathbf{x}^T\mathbf{x} & \mathbf{x}^T\mathbf{y} \\ \mathbf{y}^T\mathbf{x} & \mathbf{y}^T\mathbf{y} \end{bmatrix} = \begin{bmatrix} 1 & R \\ R & 1 \end{bmatrix}. \tag{9}$$

The spatial correlation matrix $\mathbf{C}$ can be expressed as

$$\mathbf{C} = \mathbf{X}^T\mathbf{W}\mathbf{X} = \begin{bmatrix} I_x & I_{xy} \\ I_{yx} & I_y \end{bmatrix}. \tag{10}$$

The spatial correlation matrix includes both the spatial autocorrelation and cross-correlation coefficients. The diagonal elements represent spatial autocorrelation coefficients, while the elements outside the diagonal represent spatial cross-correlation coefficients.

## 2.2 Derivation of new relations for spatial autocorrelation models

Spatial autoregressive modeling can be achieved by two different ways, which are equivalent to one another in theory. One is the direct approach. All possible spatial auto-regression and lag-regression terms are introduced to a regression model. Then, unnecessary explanatory variables are eliminated by statistical test step by step. Maybe no variable is deleted. The removed variables may be conventional independent variable, or auto-regression term, or lag-regression term. Whether to eliminate a variable depends on comprehensive statistical analysis. The other is the indirect approach. This method is to introduce the auto-regression and lag-regression terms into conventional linear regression model to eliminate spatial autocorrelation of residual sequence of the original simple model. The former can be treated as a backward method, while the latter can be regarded as a forward method. In this work, the second method are employed to generate a general spatial auto-regressive model based on one conventional explanatory variable. The postulates for the initial linear regression model are as follows. First, one influence factor. The response variable *y* depends and only depends on the independent variable *x*. In other words, no other independent variable affects *y*. Second, linear relationships. The change rate of the dependent variable *y* is in constant proportion to that of the independent variable *x*. Thus we have a simple linear regression function as follows

$$y_i = bx_i + \varepsilon_i. \tag{11}$$

in which $\varepsilon_i$ denotes residuals term. For the standardized variables, the constant term is zero, and *b*=*R*, where *R* denotes the correlation coefficient. For one variable linear regression, simple correlation coefficient equals multiple correlation coefficient. The residuals are expected to follow normal distribution. That is, $\varepsilon_i$ must be a while noise. This can be judged by Moran's index. Suppose that the standard deviation of the residuals series is $\sigma_\varepsilon$. The Moran's *I* of the residuals sequence can be estimated by

$$I_e = \mathbf{e}^T\mathbf{W}\mathbf{e} = \frac{1}{\sigma_\varepsilon}\boldsymbol{\varepsilon}^T\mathbf{W}\boldsymbol{\varepsilon}, \tag{12}$$

where **e** refers to the standardized error vector, and **ε** to the original residuals vector (Chen, 2016). For the perfect fit of a conventional linear regression model to observational data, there will be no autoregressive and lag-regressive terms. In this case, the model's structure is sound and residuals sequence bears no significant autocorrelation. Residuals autocorrelation suggests possible auto-regression, which can be reflected by Moran's index or Geary's coefficient. If the $I_e$ value is significantly different from zero, we will have spatial residuals auto-regression equation

$$\varepsilon_i = a + \rho(\mathbf{W}\boldsymbol{\varepsilon})_i + u_i = a + \rho((\mathbf{W}\mathbf{y})_i - b(\mathbf{W}\mathbf{x})_i)_i + u_i, \tag{13}$$

where *a* is constant term, $u_i$ is a white noise sequence representing the residuals of the residuals auto-regression model, and $(\mathbf{W}\boldsymbol{\varepsilon})_i = (\mathbf{W}\mathbf{y})_i - b(\mathbf{W}\mathbf{x})_i$. The variables are defined as follows

$$(\mathbf{W}\mathbf{x})_i = \sum_{j=1}^{n} w_{ij}x_i, (\mathbf{W}\mathbf{y})_i = \sum_{j=1}^{n} w_{ij}y_i, (\mathbf{W}\boldsymbol{\varepsilon})_i = \sum_{j=1}^{n} w_{ij}\varepsilon_i. \tag{14}$$

in which **Wx** denotes the lag-regression term, and **Wy** refers to the auto-regression term. Substituting equation (13) into equation (11) yields

$$y_i = a + bx_i + \rho(\mathbf{W}\boldsymbol{\varepsilon})_i + u_i = a + bx_i + \beta_1(n\mathbf{W}\mathbf{y})_i - \beta_2\rho(n\mathbf{W}\mathbf{x})_i + u_i, \tag{15}$$

where $\beta_1 = -b\rho/n$ refers to lag-regression coefficient, and $\beta_2 = \rho/n$ to auto-regression coefficient. Equation (15) can be expressed as matrix form, that is

$$\mathbf{y} = a + b\mathbf{x} + \beta_1 n\mathbf{W}\mathbf{x} + \beta_2 n\mathbf{W}\mathbf{y} + \mathbf{u}. \tag{16}$$

This can regarded as general one variable spatial autoregressive model. It is actually a mixed model, comprising conventional explanatory variable (**x**), spatial lag-regression term (**Wx**), and spatial autoregressive term (**Wy**). If $\beta_1 = 0$, will have a simple spatial autoregressive model; If $\beta_2 = 0$, will have a simple spatial lag-regressive model; If $\beta_1 = 0$ and $b = 0$, will have a pure spatial autoregressive model. Based on the first modeling method, the general model, equation (16), can be directly given, but it need a strict statistical testing. According to the test results, we can determine which explanatory variables (**x**, **Wx**, **Wy**) are retained in the expression and which explanatory ones are removed from the model.

Now, a set of formulae for the regression coefficients can be derived from the general spatial auto-regression model. The formulae are based on pure theory, but they are helpful for our understanding the structure of the spatial auto-regressive model. In standard case, the constant term $a=0$. Left multiplying equation (16) by $x^T$ and $y^T$, respectively, yields

$$\mathbf{x}^T\mathbf{y} = a\mathbf{x}^T\mathbf{o} + b\mathbf{x}^T\mathbf{x} + \beta_1 n\mathbf{x}^T\mathbf{W}\mathbf{x} + \beta_2 n\mathbf{x}^T\mathbf{W}\mathbf{y} + \mathbf{x}^T\mathbf{u}, \tag{17}$$

$$\mathbf{y}^T\mathbf{y} = a\mathbf{y}^T\mathbf{o} + b\mathbf{y}^T\mathbf{x} + \beta_1 n\mathbf{y}^T\mathbf{W}\mathbf{x} + \beta_2 n\mathbf{y}^T\mathbf{W}\mathbf{y} + \mathbf{y}^T\mathbf{u}. \tag{18}$$

where $\mathbf{o} = [1,1,\ldots,1]^T$ is an ones vector, in which all the elements are 1, and $\mathbf{x}^T\mathbf{o} = \mathbf{y}^T\mathbf{o} = 0$ because that the sum of the elements in a standardized variable is zero. According to equations (3), (4), (5), (6), and (7), the results are as below

$$nR = nb + \beta_1 nI_x + \beta_2 nI_{xy}, \tag{19}$$

$$n = nbR + \beta_1 nI_{xy} + \beta_2 nI_y + n\delta. \tag{20}$$

in which $\delta$ refers to the variance of the residuals sequence $u$. Rearranging equations (19) and (20) yields

$$\beta_1 I_x + \beta_2 I_{xy} = R - b, \tag{21}$$

$$\beta_1 I_{yx} + \beta_2 I_y = 1 - Rb - \delta. \tag{22}$$

The variance of the residuals sequence can be expressed as

$$\delta = \frac{1}{n}\mathbf{y}^T\mathbf{u} = \frac{1}{n}\mathbf{y}^T(\sigma_u \mathbf{e}_u) = \sigma_u(\frac{1}{n}\mathbf{y}^T\mathbf{e}_u) = \sigma_u R_{yu} = \sigma_u^2. \tag{23}$$

where $e_u$ denotes standardized residuals, $\sigma_u$=Var($u$) refers to the standard deviation of the residuals, $R_{yu}$ is the coefficient of correlation between dependent variable $y$ and residuals $u$. The correlation coefficient $R_{yu}$ can be proved to equal the standard deviation $\sigma_u$. In the mathematical world, there is no error. Prediction residuals appears in the real world and computational world. In this case, equation (22) can be simplified to

$$\beta_1 I_{yx} + \beta_2 I_y = 1 - Rb. \tag{24}$$

Equations (21) and (22) can be expressed as matrix equations, that is

$$\begin{bmatrix} I_x & I_{xy} \\ I_{yx} & I_y \end{bmatrix} \begin{bmatrix} \beta_1 \\ \beta_2 \end{bmatrix} = \begin{bmatrix} R-b \\ 1-Rb-\sigma_u^2 \end{bmatrix}. \tag{25}$$

According to Cramer's rule, a group of determinants can be constructed as

$$O = \begin{vmatrix} R-b & I_{xy} \\ 1-Rb-\sigma_u^2 & I_y \end{vmatrix}, \quad P = \begin{vmatrix} I_x & R-b \\ I_{yx} & 1-Rb-\sigma_u^2 \end{vmatrix}, \quad Q = \begin{vmatrix} I_x & I_{xy} \\ I_{yx} & I_y \end{vmatrix}. \tag{26}$$

The lag regression coefficient and auto-regression coefficient can be solved as below:

$$\begin{cases} \beta_1 = \dfrac{O}{Q} = \dfrac{(R-b)I_y - (1-Rb-\sigma_u^2)I_{xy}}{I_x I_y - I_{xy}^2} \\ \beta_2 = \dfrac{P}{Q} = \dfrac{(1-Rb-\sigma_u^2)I_x - (R-b)I_{xy}}{I_x I_y - I_{xy}^2} \end{cases}. \tag{27}$$

Apparently, the spatial auto- lag-regressive coefficient depend on not only spatial autocorrelation coefficients, but also spatial cross-correlation coefficient as well as Pearson correlation coefficient. The mathematical structure of the two formulates lends further support to the judgment that Moran's index bears its limitations in estimating spatial dependence (Li *et al*, 2007).

For the conventional linear regressive model based on standardized variables, the constant term is zero. However, in spatial autoregressive model, the constant term is hardly eliminated. The formula of the constant term can be derived for reference. Summing equation (15) yields

$$\sum_{i=1}^{n} y_i = \sum_{i=1}^{n} a + b\sum_{i=1}^{n} x_i + \beta_1 n \sum_{i=1}^{n} (\mathbf{Wx})_i + \beta_2 n \sum_{i=1}^{n} (\mathbf{Wy})_i + \sum_{i=1}^{n} u_i, \tag{28}$$

where

$$\sum_{i=1}^{n} y_i = 0, \sum_{i=1}^{n} a = na, \sum_{i=1}^{n} x_i = 0, \sum_{i=1}^{n} u_i = 0. \tag{29}$$

Thus we have

$$a = -\beta_1 \sum_{i=1}^{n} (\mathbf{Wx})_i - \beta_2 \sum_{i=1}^{n} (\mathbf{Wy})_i = -\beta_1 \mathrm{E}((n\mathbf{Wx})_i) - \beta_1 \mathrm{E}((n\mathbf{Wy})_i). \tag{30}$$

where $\mathrm{E}(\bullet)$ implies averaging the elements of a variable. For the ordinary linear regression model based on one independent variable, the regression coefficient is

$$b = \frac{\sum_{i=1}^{n}(x_i - \bar{x})(y_i - \bar{y})}{\sum_{i=1}^{n}(x_i - \bar{x})^2} = \frac{\mathrm{cov}(x,y)}{\mathrm{Var}(x)} = \frac{\mathrm{cov}(x,y)}{\sigma_x \sigma_y} \frac{\sigma_y}{\sigma_x} = R\frac{\sigma_y}{\sigma_x}, \tag{31}$$

where $\mathrm{cov}(x,y)$ denotes the covariance between $x$ and $y$, $\mathrm{Var}(x) = \sigma_x^2$ is the population variance of $x$, and $\sigma_x$ and $\sigma_y$ represent the standard deviations of $x$ and $y$, respectively. Since the variables are standardized, the standard deviations are equal to 1, that is, $\sigma_x = \sigma_y = 1$. Therefore, we have

$$b = R = \frac{\mathrm{cov}(x,y)}{\sigma_x \sigma_y}. \tag{32}$$

In a theoretical derivation process, the error item can be ignored, and thus we have $\delta = 0$. In this instance, equation (27) can be simplified as

$$\begin{cases} \beta_1 = \dfrac{(R-b)I_y - (1-Rb)I_{xy}}{I_x I_y - I_{xy}^2} \\ \beta_2 = \dfrac{(1-Rb)I_x - (R-b)I_{xy}}{I_x I_y - I_{xy}^2} \end{cases}. \tag{33}$$

If $b=R$ as indicated above, then equation (33) can be further simplified as

$$\begin{cases} \beta_1 = \dfrac{(R^2-1)I_{xy}}{I_x I_y - I_{xy}^2} \\ \beta_2 = \dfrac{(1-R^2)I_x}{I_x I_y - I_{xy}^2} \end{cases}. \tag{34}$$

If $I_{xy}=0$, but $I_x \neq 0$, $I_y \neq 0$, $I_x I_y \neq I_{xy}^2 = 0$, we have a simple spatial auto-regressive model as follows

$$\mathbf{y} = a + b\mathbf{x} + \beta_2(n\mathbf{Wy}) + \mathbf{u}. \tag{35}$$

No lag-regression term in equation (35), which can be given by analyzing spatial errors (Ward and Gleditsch, 2008). If $b=0$ in equation (36), we will have a pure spatial autoregressive model (Li *et al*, 2007; Odland, 1988). The pure spatial autoregressive model based on standardized variable can be derived from one of canonical equation for Moran's index, $I\mathbf{z}=n\mathbf{Wz}+\mathbf{u}$, in which $a=0$ (Chen, 2013). So we have $\beta_2=1/I_y$ in theory. If $I_x=0$, $I_{xy}\neq 0$, $I_xI_y\neq I_{xy}^2$, we have simple spatial lag-regressive model as below

$$\mathbf{y} = a + b\mathbf{x} + \beta_1(n\mathbf{Wx}) + \mathbf{u}. \tag{36}$$

No auto-regression term in equation (36). Equations (35) and (36) are special cases of equation (16). If $b=0$ in equation (36), we will have a pure spatial lag-regressive model. In this case, $a=0$, and we have $\beta_1=R/I_x$ in theory.

## 2.3 Derivation of collinearity criterion

The spatial auto-regressive model is actually a linear multivariable model. Maybe there are collinear relationships between different explanatory variables. Collinearity of explanatory variables may lead to abnormal or even absurd estimated values of regressive coefficients. The most possible collinearity relationship comes between the auto-regressive term ($\mathbf{Wy}$) and lag-regressive term ($\mathbf{Wx}$). The linear relation can be expressed as

$$\mathbf{Wy} = c + d\mathbf{Wx}, \tag{37}$$

where $c$ and $d$ are two constants. On the other hand, according to equation (26), we have

$$Q = \begin{vmatrix} I_x & I_{xy} \\ I_{yx} & I_y \end{vmatrix} = I_x I_x - I_{xy}^2. \tag{38}$$

Left multiplying equation (37) by $\mathbf{x}^T$ and $\mathbf{y}^T$, respectively, yields

$$\mathbf{x}^T\mathbf{Wy} = c\mathbf{x}^T\mathbf{o} + d\mathbf{x}^T\mathbf{Wx}, \tag{39}$$

$$\mathbf{y}^T\mathbf{Wy} = c\mathbf{y}^T\mathbf{o} + d\mathbf{y}^T\mathbf{Wx}. \tag{40}$$

As indicated above, $\mathbf{x}^T\mathbf{o}=0$, $\mathbf{y}^T\mathbf{o}=0$. In terms of equations (5), (6), and (7) shown above, equations (39) and (40) can be expressed as

$$I_{xy} = dI_x, \tag{41}$$

$$I_y = dI_{yx}. \tag{42}$$

Multiplying the above two equations one another yields

$$I_x I_y = I_{xy}^2, \tag{43}$$

which gives the theoretical condition of collinearity between auto-regressive term (**Wy**) and lag-regressive term (**Wx**), that is, the determinant $Q=0$. This suggest that if **Wy** and **Wx** are of collinear relation, the lag-regressive coefficient ($\beta_2$) and auto-regressive coefficient ($\beta_1$) will become meaningless. To avoid collinearity, quantitative geographers and spatial statisticians often directly discard spatial lag-regressive term, **Wx**, and take a spatial autoregressive model with form similar to equation (35) (Anselin, 1988; Ward and Gleditsch, 2008).

## 3 Empirical analysis

### 3.1 Study area, data, and algorithm

The effect of mathematical derivation needs to be evaluated by observational data. If a reasoning result is correct, it will be consistent with the calculated results based on observed data. In fact, the success of sciences rest with their great emphasis on the role of quantifiable data and their interplay with models (Louf and Barthelemy, 2014). In this section, several sets of observational data will be employed to testify the derived results above. The study area is Beijing-Tianjin-Hebei (BTH) region of China (Figure 1). It is also termed Jing-Jin-Ji (JJJ) region in literature. There are three sources of observational data. The spatial distances are measured by traffic mileage, which were extracted by ArcGIS. City sizes were measured by urban nighttime light (NTL) area and urban population (Table 1). The data of NTL area are defined within built-up area of cities in BTH region (Chen and Long, 2021; Long and Chen, 2019). The data of urban population, including city population and town population, come from census in 2000 (the fifth census) and 2010 (the sixth census). The spatial proximity is defined by $v_{ij}=1/r_{ij}$, where $r_{ij}$ denotes the traffic mileage between city $i$ and city $j$. Thus the spatial contiguity matrix can be expressed as $\mathbf{V}=[v_{ij}]=[1/r_{ij}]$, in which the diagonal elements are defined as zero. Normalizing **V** yields a spatial weight matrix **W**, and the summation of the elements in **W** is 1. In light of the idea from allometric scaling relation, urban population and NTL area are taken natural logarithms, that is, let $x$=ln(urban population) and $y$=ln(NTL area).

**Table 1 The measures and data sources for empirical analysis of normalized spatial auto-regression**

| Measure | Symbol | Meaning | Data source | Year |
|---|---|---|---|---|
| **Distance** | $r_{ij}$ | Interurban distance | Extraction by ArcGIS | 2010 |
| **City size 1** | $x_i$ | Natural logarithm of urban population (city and town population) | The fifth and sixth census of China | 2000, 2010 |
| **City size 2** | $y_i$ | Natural logarithm of Nighttime light (NTL) area | American NOAA National Centers for Environmental Information (NCEI) | 2000, 2010 |

**Note**: NTL data were processed by Long and Chen (2019) and Chen and Long (2021).

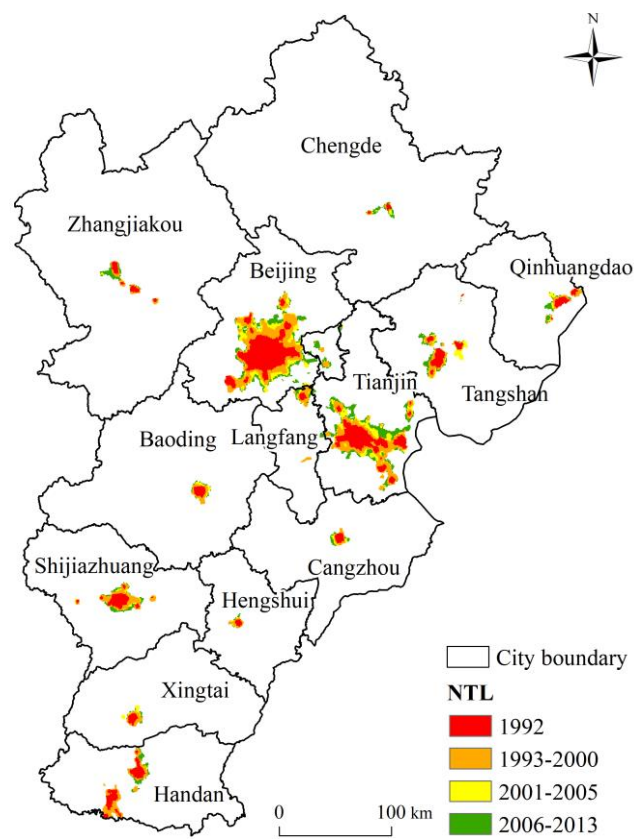

**Figure 1 The system of cities in Beijing, Tianjin, and Hebei region, China**

(**Note**: The city sizes and their changes are illustrated by the nighttime light strength and area.)

The multivariate linear regression based on ordinary least squares (OLS) method is a classic framework for estimating autoregressive parameter values. Based on matrixes and vectors, regressive coefficients can be calculated by following formula (Cliff and Ord, 1981; Permai *et al*, 2019)

$$\mathbf{b} = (\mathbf{X}^T\mathbf{X})^{-1}\mathbf{X}^T\mathbf{y}, \tag{44}$$

in which *b* refers to the vector of the parameters of spatial autoregressive models, equations (16), (35), and (36). The meanings of other symbols are the same as those in equations (1), (2), and (8). For equation (16), we have

$$\mathbf{b} = \begin{bmatrix} a & b & \beta_1 & \beta_2 \end{bmatrix}^T. \tag{45}$$

The rest may be deduced by analogy. For convenience, multiple linear regression technique based on LSM may be used to estimate the model parameters for empirical analysis.

## 3.2 Results and analysis

If we take the major cities in the study area into account, we will have a small spatial sample consisting of 13 elements. The 13 cities were certified as prefecture level cities in BTH region according to the urban standard of China. As far as empirical research is concerned, the results based on small samples are not convincing in many cases. Where a teaching case is concerned, the results based on small samples are concise and easy to understand. The goal of this paper is at theoretical research rather than empirical analysis. Therefore, a simple case is helpful for readers to understand the application of the newly derived formulae. The processes of calculation and analysis are fulfilled step by step as follows.

**Step 1, preliminary calculation**. First, based on mean and population standard deviation, the two size variables, that is, urban population (**x**) and nighttime light (NTL) area (**y**), can be standardized according to *z*-score formula. The spatial proximity matrix (**V**) can be normalized so that it become a spatial weight matrix (**W**). Then, the Pearson correlation coefficient (*R*) can be computed by using equation (4), two spatial autocorrelation coefficients, i.e., Moran's indexes, can be calculated for population and NTL area by using equation (5) and (6), the coefficient of spatial cross-correlation between population and NTL area can be calculated by using equation (7). In fact, using equation (10), we can compute both spatial auto-correlation and cross-correlation coefficients meantime conveniently. For example, for 2010, the results are as below:

$$\hat{\mathbf{C}} = \mathbf{X}^T\mathbf{W}\mathbf{X} = \begin{bmatrix} -0.1812 & -0.1287 \\ -0.1287 & -0.0694 \end{bmatrix}, \tag{46}$$

where the symbol "^" indicates estimated value. Taking **x** and **y** as independent variables,

respectively, and $n\mathbf{W}\mathbf{x}$ and $n\mathbf{W}\mathbf{y}$ as dependent variables, respectively, we can estimate the spatial auto-correlation and cross-correlation coefficients and give the *P*-values indicating significance levels by ordinary one-variable linear regression analysis (Table 2).

**Table 2 Preparing and complementary calculated results: spatial correlation indexes, Pearson correlation coefficient, and residuals variances**

| Type | Correlation parameter | Results for 2000 | | Results for 2010 | |
|---|---|---|---|---|---|
| | | Index | *P*-value | Index | *P*-value |
| **Spatial correlation** | $I_x$ | -0.1940 | 0.0273 | -0.1812 | 0.0455 |
| | $I_{xy}$ | -0.1459 | 0.1183 | -0.1287 | 0.1911 |
| | $I_{yx}$ | -0.1459 | 0.1141 | -0.1287 | 0.1747 |
| | $I_y$ | -0.0968 | 0.3145 | -0.0694 | 0.4926 |
| **Pearson correlation** | $R$ | 0.9571 | | 0.9534 | |
| **Residuals variance** | $\sigma^2$ | 0.0575 | | 0.0583 | |

**Step 2, parameter estimation**. First, we can estimate the parameter values of the spatial autoregressive models by means of theoretical formulae, equations (30) and (34). Second, using multivariable linear regressive analysis based on the OLS method, we can estimate the model parameters from the empirical viewpoint (Table 3). Empirical calculation can be carried out with the help of spreadsheet, Microsoft Excel, or mathematical calculation software, or statistical analysis software. It is necessary to make more about the direct calculation method based on formulae derived in Section 2. In theory, $b=R$, $\sigma_u^2=0$. Thus, we can use equation (34) to calculate the autoregressive coefficients, $\beta_1$ and $\beta_2$, and then use equation (30) to work out the constant term, *a*. For example, for 2010, the results of the autoregressive coefficients are as follows

$$\begin{cases} \hat{\beta}_1 = \dfrac{(0.9534^2 - 1)*(-0.1287)}{(-0.1812)*(-0.0694)-(-0.1287)^2} = -2.9388 \\ \hat{\beta}_2 = \dfrac{(1-0.9534^2)*(-0.1812)}{(-0.1812)*(-0.0694)-(-0.1287)^2} = 4.1392 \end{cases}. \tag{47}$$

The means of $n\mathbf{W}\mathbf{x}$ and $n\mathbf{W}\mathbf{y}$ are 0.1137 and 0.1256, respectively. So, in light of equation (30), the constant term is

$$\hat{a} = 2.9388*0.1137 - 4.1392*0.1256 = -0.1858. \tag{48}$$

The calculation process is based on 14 digits after the decimal point, and only 4 digits after the decimal point are displayed in equations (46) and (47), so there are slight errors in the numbers on both sides of the equal signs.

Table 3 Empirical and theoretical results of spatial auto- and lag-regression model parameter estimation

| Type | Parameter | Results for 2000 | | | Results for 2010 | | |
|---|---|---|---|---|---|---|---|
| | | Theoretical | Empirical | | Theoretical | Empirical | |
| | | Parameter value | Parameter value | $P$-value | Parameter value | Parameter value | $P$-value |
| General spatial autoregressive model | $a$ | -0.3334 | -0.0724 | 0.4416 | -0.1858 | -0.0740 | 0.4220 |
| | $b$ | 0.9571 | 1.0861 | 0.0000 | 0.9534 | 1.0087 | 0.0000 |
| | $\beta_1$ | -4.9221 | 0.6637 | 0.6603 | -2.9388 | -0.3170 | 0.8264 |
| | $\beta_2$ | 6.5455 | 0.0021 | 0.9988 | 4.1392 | 0.8763 | 0.4907 |
| Spatial Autoregressive model | $a$ | -0.8686 | -0.0781 | 0.3823 | -0.5199 | -0.0764 | 0.3802 |
| | $b$ | 0.9571 | 1.0430 | 0.0000 | 0.9534 | 1.0316 | 0.0000 |
| | $\beta_2$ | 6.5455 | 0.5888 | 0.0655 | 4.1392 | 0.6078 | 0.0406 |
| Spatial lag-regressive model | $a$ | 0.5352 | -0.0724 | 0.4035 | 0.3341 | -0.0757 | 0.3982 |
| | $b$ | 0.9571 | 1.0862 | 0.0000 | 0.9534 | 1.0741 | 0.0000 |
| | $\beta_1$ | -4.9221 | 0.6659 | 0.0574 | -2.9388 | 0.6662 | 0.0539 |

**Step 3, checking computations**. The key equations and formulae can be testified by the calculation result shown above. As a preparation, it is necessary to calculate the variances of the model prediction residuals ($\sigma_u^2$) by using equation (23) (see Table 2 for the results). Then, we can verify equations (21) and (24) by using the results based on theoretical formulae, and verify equations (21) and (22) by using the results based on least squares regression. For example, for 2010, according to the theoretical formulae, we have

$$\beta_1 I_x + \beta_2 I_{xy} = -2.9388*(-0.1812)+4.1392*(-0.1287)=0, \qquad (49)$$

$$R - b = 0.9534 - 0.9534 = 0. \qquad (50)$$

The left term equals the right term. According to the empirical analysis, we have

$$\beta_1 I_x + \beta_2 I_{xy} = -0.3170*(-0.1812)+0.8763*(-0.1287)=0.0553, \qquad (51)$$

$$R - b = 0.9534 - 1.0087 = 0.0553. \qquad (52)$$

The left term is still equal to the right term. The rest can be verified by the similar way (Table 4). Further, we can testify equation (27). For example, for 2010, $b$=1.0087, $R$=0.9534, $\sigma_u^2$=0.0583. $I_x$=-0.1812, $I_{xy}$=-0.1287, $I_y$=-0.0694 (Tables 2 and 3), substituting these results into equation (27) yields $\beta_1$=-0.3170, $\beta_2$=0.8763. As indicated above, the means of $n\mathbf{Wx}$ and $n\mathbf{Wy}$ are 0.1137 and 0.1256, respectively. Substituting these results into equation (30) yields $a$=-0.0740. These results are the same as those from the least squares calculation.

**Table 4 Checking calculations for the typical equations for the relationships between spatial autocorrelation and spatial auto-regression analysis**

| Type | Equation | Results for 2000 | | Results for 2010 | |
| --- | --- | --- | --- | --- | --- |
| | | Left | Right | Left | Right |
| **Theoretical** | Equation (21) | 0 | 0 | 0 | 0 |
| | Equation (24) | 0.0840 | 0.0840 | 0.0910 | 0.0910 |
| **Empirical** | Equation (21) | -0.1290 | -0.1290 | -0.0553 | -0.0553 |
| | Equation (22) | -0.0970 | -0.0970 | -0.0200 | -0.0200 |

The formulae derived above is for helping geographers understand the relationships between spatial correlation and spatial auto-regression rather than providing them with a new algorithm. The autoregressive term and lag-regressive term cannot be introduced into the models of Beijing-Tianjin-Hebei urban system of 13 cities meantime because of collinearity (Figure 2). One of them must be discarded, otherwise the confidence level of parameter estimation values is too low (see Table 3). However, it is not easy to judge which term to retain, spatial autoregressive term or lag regression term. Using the formulae, we can make a judgement in terms of the results of spatial correlation coefficients (see Table 2). Suppose that the significance level is taken as 0.05. The Moran's index of urban population size (independent variable $x$) is significant, but the Moran's index of NTL area (dependent variable $y$) is not significant. The coefficients of cross-correlation between urban population and NTL area is not so significant. In this case, according to equation (33) or (34), we keep the spatial autoregressive term but discard the spatial lag-regressive term. Thus, we have a spatial autoregressive model for 2000 as follows

$$\hat{y}_i = -0.0781 + 1.0430 x_i + 0.5888 \sum_{j=1}^{n} w_{ij} y_i + u_i. \tag{53}$$

As for 2010, the model is as below

$$\hat{y}_i = -0.0764 + 1.0316 x_i + 0.6078 \sum_{j=1}^{n} w_{ij} y_i + u_i. \tag{54}$$

The basic statistics for testing the models are tabulated for reference (Table 5). It can be seen that the quality of the model has improved from 2000 to 2010. This option is not particularly satisfactory because that the confidence level of Moran's index of NTL area (dependent variable $y$) is low.

However, comprehensive analysis show that the choice of spatial autoregressive model is better than the choice of spatial lag regression model.

**Table 5 The basic statistics for testifying the spatial autoregressive models from global and local levels**

| Type | Statistic/Parameter | Results for 2000 | Results for 2010 |
|---|---|---|---|
| **Global test** | Goodness of fit $R^2$ | $R^2$=0.9412 | $R^2$=0.9414 |
| | Regression standard error $s$ | $s$=0.2766 | $s$=0.2761 |
| | $F$ statistic $F$ | $F$=79.9862 | $F$=80.2963 |
| | Spatial Durbin-Watson statistics | $DW$=1.5615 | $DW$=1.7718 |
| **Local test** | Constant term $a$ | $P$=0.3823 | $P$=0.3802 |
| | Conventional regressive coefficient $b$ | $P$=0.0000 | $P$=0.0000 |
| | Autoregressive coefficient $\beta_2$ | $P$=0.0655 | $P$=0.0406 |

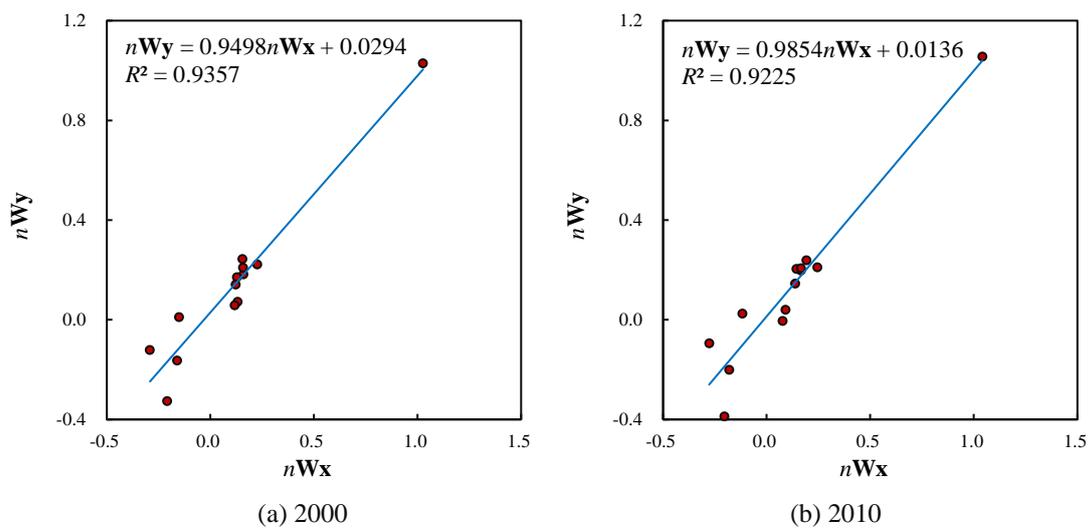

(a) 2000  (b) 2010

**Figure 2 The patterns of collinear relationships between spatial autoregressive term and spatial lag-regressive term**

(**Note**: For normalization, the autoregressive term **Wy** and lag-regressive term **Wx** were multiplied by $n$.)

## 4 Discussion

Moran's index proceeded from Pearson's product-moment correlation coefficient. Generalizing Pearson's simple correlation coefficient to a 1-dimensional autocorrelation coefficient of time series analysis, and then, by analogy, substituting time lag parameter with spatial displacement parameter, we will have a 1-dimension spatial autocorrelation coefficient. Further, replacing the displacement parameter based on 1-dimensional space with a spatial contiguity matrix based on 2-dimensioanl space, we will have a 2-dimensional spatial autocorrelation coefficient, i.e., Moran's *I*. Finally, by analogy with Moran's index and Pearson correlation coefficient, we have spatial cross-correlation coefficient (Table 6). Moran's index is a spatial statistic used to measure the spatial dependence of different sampling points in a geographical region. The precondition of statistical analysis validity of a spatial sample based on conventional quantitative method is independence of sampling points (Haggett *et al*, 1977; Odland, 1988). Moran's index can be used to judge whether or not a sampling point is independent of other sampling points in a spatial sample. If there is no significant correlation between different sampling points, we will be able to make effective parameter modeling and quantitative analysis using conventional statistical methods such as regression analysis, principal component analysis, and cluster analysis. In contrast, if there is significant correlation between these sampling points, conventional statistical methods will fail, and we'd better make spatial autocorrelation analysis. Based on spatial autocorrelation, spatial auto-regressive modeling can be made for geographical systems. It can be seen that spatial autocorrelation measurement was initially used to reveal problems, but later was developed into an analysis method.

**Table 6 Similarities and differences between correlation coefficient, autocorrelation function, and spatial autocorrelation coefficient**

| Type | Property | Object | Formula |
|---|---|---|---|
| Pearson's product-moment correlation coefficient | Simple cross correlation | Two random variables for time series, spatial series, or random data points (two samples) | $R = \frac{1}{n}\mathbf{x}^T\mathbf{y} = \frac{1}{n}\mathbf{y}^T\mathbf{x}$ |
| Autocorrelation function for time series | One-dimensional temporal autocorrelation | Ordered time series (one sample) | $\rho_\tau = \mathbf{x}^T\mathbf{W}_\tau\mathbf{x} = \frac{1}{2n}\mathbf{x}^T\mathbf{V}_\tau\mathbf{x}$ |
| Autocorrelation function for spatial series | One-dimensional spatial autocorrelation | Ordered spatial series (one sample) | $\rho_k = \mathbf{x}^T\mathbf{W}_k\mathbf{x} = \frac{1}{2n}\mathbf{x}^T\mathbf{V}_k\mathbf{x}$ |
| Moran's index | Two-dimensional spatial autocorrelation | Random spatial observational points (one sample) | $I = \mathbf{x}^T\mathbf{W}\mathbf{x}$ |
| Cross-correlation coefficient | Two-dimensional spatial cross- | Random spatial observational points | $I_{xy} = \mathbf{x}^T\mathbf{W}\mathbf{y} = \mathbf{y}^T\mathbf{W}\mathbf{x}$ |

| | correlation | (two samples) | |

**Note**: **V** denotes spatial contiguity matrix, and **W** refers to spatial weight matrix, $\rho$ is autocorrelation function, $\tau$ is time lag parameter, and $k$ is spatial displacement parameter, which is always regarded as spatial lag in literature. As for the matrix expression of temporal autocorrelation coefficient, see Appendix.

Spatial autocorrelation is the precondition of spatial autoregressive modeling. However, how to understand autoregressive coefficient through autocorrelation coefficient is an outstanding problem to be solved for a long time. As an attempt, the mathematical structure of spatial auto-regression coefficients based on standardized variables were decomposed into a set of statistic parameters in this work. These statistic parameters include Pearson correlation coefficient ($R$), ordinary regression coefficient ($b$), spatial autocorrelation coefficients (Moran's indexes $I_x$ and $I_y$), and spatial cross-correlation coefficient ($I_{xy}$). Thus, the relationships between ordinary linear regression, spatial auto- and cross-correlation, and spatial auto-regression are revealed exactly. The results are helpful for deep understanding the inherent relationships between spatial correlation analysis and spatial autoregressive modeling. Generally speaking, if all the autocorrelation coefficient ($I_x$ and $I_y$) and cross-correlation coefficient of the independent and dependent variables ($I_{xy}$) are significant, and the product of two autocorrelation coefficients is not equal to the square of the cross-correlation coefficient, we will have a general spatial autoregressive model including both spatial autoregressive term and lag-regressive term (**Wx**, **Wy**). Three typical special cases are summarized as follows. (1) If all the autocorrelation coefficient of the independent variable ($I_x$) and cross-correlation coefficient ($I_{xy}$) are significant, and the product of two autocorrelation coefficients is not equal to the square of the cross-correlation coefficient, we will still have a general spatial autoregressive model including both spatial autoregressive term and lag-regressive term (**Wx**, **Wy**); (2) If all the autocorrelation coefficient ($I_x$ and $I_y$) of the independent and dependent variables ($I_{xy}$) are significant, but the cross-correlation coefficient of two variables are not significant, we will have a special spatial autoregressive model including only spatial autoregressive term (**Wy**) but no lag-regressive term (**Wx**); (3) If the cross-correlation coefficient of two variables are significant, but the autocorrelation coefficient of the independent variable ($I_x$) is not significant, we will have a special lag-autoregressive model including only spatial lag-regressive term (**Wx**) but no autoregressive term (**Wy**) (Table 7).

**Table 7 Three types of simple spatial autoregressive and lag-regressive models and corresponding**

**correlation conditions in theory**

| Type | Correlation condition | Regression model | Regression coefficient |
|---|---|---|---|
| General spatial auto-regression | $I_x \neq 0$<br>$I_{xy} \neq 0$<br>$I_x I_y \neq I_{xy}^2$ | $\mathbf{y} = a + b\mathbf{x} + \beta_1 n\mathbf{Wx} + \beta_2 n\mathbf{Wy} + \mathbf{u}$ | $\begin{cases} \beta_1 = \dfrac{(R^2-1)I_{xy}}{I_x I_y - I_{xy}^2} \\ \beta_2 = \dfrac{(1-R^2)I_x}{I_x I_y - I_{xy}^2} \end{cases}$ |
| Special spatial auto-regression | $I_x \neq 0$<br>$I_y \neq 0$<br>$I_{xy} = 0$ | $\mathbf{y} = a + b\mathbf{x} + \beta_2 n\mathbf{Wy} + \mathbf{u}$ | $\beta_2 = \dfrac{1-R^2}{I_y}$ |
| Spatial lag-regression | $I_x = 0$<br>$I_{xy} \neq 0$ | $\mathbf{y} = a + b\mathbf{x} + \beta_1 n\mathbf{Wx} + \mathbf{u}$ | $\beta_1 = \dfrac{R^2-1}{I_{xy}}$ |

Unfortunately, in empirical analysis, things become more complicated than those in the theoretical world. Spatial correlation process is actually a spatio-temporal feedback process. A series of feedback loops involve both time lags and spatial displacements, which result in response delay of spatial interaction. Response delay leads to nonlinearity, which in turn lead to spatial complexity. One of the troubles may lie in the paradox of spatial autocorrelation measurement. As we know, spatial autocorrelation coefficients are calculated by average values and corresponding standard deviation, and a standard deviation is based on an average value. This suggests that the reliability of the average values is the premise of whether the calculation results of spatial autocorrelation coefficients is reliable. The basic average value is the arithmetic mean, which is obtained by dividing the sum by the total number. If and only if there is no significant spatial autocorrelation causing affine correlation and information redundancy, the digital information of the sum in a geographical region is equal to the digital information of the sum of parts. Otherwise, the digital information of the sum is unequal to the digital information of the sum of parts, and the average value is not so valid. This implies that if the spatial autocorrelation coefficient is not significant, the result is credible; on the contrary, if the spatial autocorrelation is significant, the autocorrelation coefficient value is not so reliable. This is a quasi-paradox or semi-paradox, which can be called "spatial autocorrelation paradox". Owing to this paradox, it is difficult to make a simple judgment for spatial autoregressive model selection by spatial a set of correlation coefficients (autocorrelation coefficients and cross-correlation coefficients). For the time being, the only way to solve the problem is to make a comprehensive judgment by integrating the formula in this paper and the statistical test of regression analysis.

As indicated above, spatial autocorrelation naturally lead to spatial autoregressive modeling. Spatial autoregressive research can be traced back to the early years of biometrics (Whittle, 1954). But it is quantitative and theoretical geographers as well as econometricians who developed the theory and method of spatial autoregressive analysis systematically (Anselin, 1988; Bennet, 1979; Cliff and Ord, 1981; Griffith, 2003; Haining, 1979; Haining, 1980; LeSage, 1997; Odland, 1988; Pace and Barry, 1997; Upton and Fingleton, 1985). In literature, the spatial autoregressive models were made by taken spatial lag and spatial errors from two aspects. Generally speaking, spatial autoregressive term was introduce into a linear function, and then revise the model by analyzing spatial error autocorrelation (Haggett *et al*, 1977; LeSage, 2000; Permai *et al*, 2019; Ward and Gleditsch, 2008). In this paper, spatial lags and spatial errors were treated as the different sides of the same coin. Compared with previous studies on spatial autoregressive modeling, the novelty of this work lies in three aspects. First, based on standardized variables, Moran's index, spatial cross-correlation coefficient are associated with spatial autoregressive coefficients. Both spatial correlation and spatial auto-regression are integrated into a simple logic framework. The new framework may lead to further studies on spatial modeling and geographical analysis. Second, new formulae are derived for spatial autoregressive models. As indicated above, through this mathematical relations, we can obtain a new understanding of the conditions of spatial autoregressive modeling. Third, a study case of spatial analysis was provided about Chinese cities. The basic features of spatial correlation pattern and process of Chinese cities were illustrated by this example. What is more, several concepts are clarified. Time lag differs from spatial displacement despite the inherent relation between them, and, in the sense of spatial statistics, spatial autoregressive terms is different from spatial lag-regressive term.

The formulae of spatial autoregressive coefficients are rigorously derived from spatial regressive model and correlation parameters, and the results can stand the verification of observation data. The main shortcomings of this study mainly lies in two aspects. First, except for the autoregressive term and lag-regression term, only one ordinary explanatory variable is considered in the model. In other words, the spatial auto-regression model is derived from a one variable ordinary linear regression model by revising the residuals term. Second, the cross term based on spatial weight has not been taken into consideration in the modeling process. Spatial cross-correlation suggests possible cross term in the spatial auto-regression model. If so, the model is involved to nonlinear structure. The one variable spatial auto-regression model can be easily generalized to multiple variable spatial auto-regression model. In addition, where algorithm is concerned, the maximum likelihood estimator (MSE) is not taken into account for the time being.

## 5 Conclusions

Changing the expression forms of a mathematical model or formula can sometimes bring unexpected results. The results of theoretical derivation are based on the variable relationships defined in the mathematical world. If and only if the variables meet a set of strict conditions, the formulae can be directly employed to estimate autoregressive and lag-regressive coefficients for the spatial autoregressive models. Nevertheless, these formulas bring us a new understanding of spatial autoregressive modeling. Based on the newly derived formulae and empirical analysis, at least two points of new knowledge can be seen as follows. First, significant spatial autocorrelation of the independent variable ($x$) is the most necessary condition for introducing spatial auto-regression term into the model. This is different from our intuitive understanding. According to our intuitive understanding, the significant spatial autocorrelation of the dependent variable ($y$) is a necessary condition for considering spatial auto-regression term. Second, significant spatial cross-correlation between independent variable ($x$) and dependent variable ($y$) is the most necessary condition for spatial introducing the lag-regression term into the model. This is also different from our intuitive understanding. In light of intuitive understanding, significant spatial autocorrelation of independent variables ($x$) is the necessary condition for taking spatial lag regression term into account. Third, for the case of one ordinary explanatory variable, the conditions of autoregressive modeling can be summarized as follows. (1) If the autocorrelation coefficient of the independent variable ($I_x$) and the cross-correlation coefficient of the independent and dependent variables ($I_{xy}$) are significant, and the product of two autocorrelation coefficients is not equal to the square of the cross-correlation coefficient ($I_xI_y \neq I_{xy}^2$)), we should adopt the general spatial autoregressive model contain the conventional term (**x**), the autoregressive term (**Wy**), and lag-regressive terms (**Wx**). (2) If both the two autocorrelation coefficients of the independent and dependent variables ($I_x$ and $I_y$) are significant, but the cross-correlation coefficient of the two variables ($I_{xy}$) is not significant, we should choose the special spatial autoregressive model containing only the conventional term (**x**) and the autoregressive term (**Wy**). (3) If the cross-correlation coefficient of the independent and dependent variables ($I_{xy}$) are significant, but the autocorrelation coefficient of the independent variable ($I_x$) is not significant, we should select the special spatial lag-regressive model containing only the conventional term (**x**) and the lag-regressive term (**Wx**).

**Acknowledgement:**

This research was sponsored by the National Natural Science Foundation of China (Grant No. 42171192). The support is gratefully acknowledged.

## Appendix: Matrix expression of time series autocorrelation function

Suppose a sample path with length *n* extracting from a time series. The sample path has been standardized by means of *z*-score based on population standard deviation. It can expressed as below:

$$\mathbf{x} = \begin{bmatrix} x_1 & x_2 & \cdots & x_n \end{bmatrix}^{\mathrm{T}}. \tag{A1}$$

The sample autocorrelation coefficient can be expressed as

$$\rho_\tau = \frac{\sum_{t=\tau+1}^{n} [(x_t - \overline{x})(x_{t-\tau} - \overline{x})]}{\sum_{t=1}^{n} (x_t - \overline{x})^2}, \tag{A2}$$

in which *t*=1,2,…, *n* is time order number, *τ*=1,2,…,*t* denote time lag. Equation (A2) can be re-expressed as matrix form. Based on step function, spatial contiguity can be defined as

$$v_{ij} = \begin{cases} 1, |i-j| = 1 \\ 0, |i-j| \neq 1 \end{cases}, \tag{A3}$$

where *i, j*=1,2,…,*n*. For time lag *τ*=1, the spatial contiguity matrix can be expressed as

$$\mathbf{V}_1 = \left[v_{ij}\right]_{n \times n} = \begin{bmatrix} 0 & 1 & 0 & \cdots & 0 \\ 1 & 0 & 1 & \cdots & 0 \\ 0 & 1 & 0 & \cdots & 0 \\ \cdots & \cdots & \cdots & \ddots & \cdots \\ 0 & 0 & 0 & \cdots & 0 \end{bmatrix}. \tag{A4}$$

A global normalized spatial weight matrix can be defined as

$$\mathbf{W} = \frac{\mathbf{V}}{V_0} = \frac{1}{2n}\mathbf{V}, \tag{A5}$$

where $V_0 = 2n$. Thus we have 1-order autocorrelation coefficient

$$\rho_1 = \mathbf{x}^T \mathbf{W}_1 \mathbf{x} = \frac{1}{2n}\mathbf{x}^T \mathbf{V}_1 \mathbf{x}. \tag{A6}$$

For time lag $\tau=2$, the spatial contiguity matrix can be expressed as

$$\mathbf{V}_2 = \left[v_{ij}\right]_{n \times n} = \begin{bmatrix} 0 & 0 & 1 & \cdots & 0 \\ 0 & 0 & 0 & \cdots & 0 \\ 1 & 0 & 0 & \cdots & 0 \\ \cdots & \cdots & \cdots & \ddots & \cdots \\ 0 & 0 & 0 & \cdots & 0 \end{bmatrix}. \tag{A7}$$

Thus we have 2-order autocorrelation coefficient

$$\rho_2 = \mathbf{x}^T \mathbf{W}_2 \mathbf{x} = \frac{1}{2n}\mathbf{x}^T \mathbf{V}_2 \mathbf{x}. \tag{A8}$$

Generally, we have an autocorrelation function as follows

$$\rho_\tau = \mathbf{x}^T \mathbf{W}_\tau \mathbf{x} = \frac{1}{2n}\mathbf{x}^T \mathbf{V}_\tau \mathbf{x}. \tag{A9}$$

It can be proved that equation (A9) is mathematically equivalent to equation (A2).